\documentclass[preprint]{elsarticle}

\usepackage{amssymb}
\usepackage{color}

\journal{Superlattices and Microstructures}

\begin{document}

\begin{frontmatter}



\title{Twisted bilayer blue phosphorene: A direct band gap semiconductor}


\author[udea]{D. A. Ospina}
\author[udea]{C. A. Duque}
\author[udem]{J. D. Correa}
\author[utfsm]{Eric Su\'arez Morell}

\address[udea]{Grupo de Materia Condensada-UdeA, Instituto de F\'{\i}sica, Facultad
de Ciencias Exactas y Naturales, Universidad de Antioquia UdeA, Calle 70 No. 52-21,Medell\'{\i}n}
\address[udem]{Departamento de Ciencias B\'{a}sicas, Universidad de  Medell\'{\i}n, Medell\'{\i}n, Colombia}
\address[utfsm]{Departamento de F\'{i}sica, Universidad T\'{e}cnica Federico Santa Mar\'{i}a, Casilla 110-V, 
Valpara\'{i}so, Chile}

\begin{abstract}
We report that two rotated layers of blue phosphorene behave as a direct band gap semiconductor. The optical 
spectrum shows absorption peaks in the visible region of the spectrum and in addition the energy of these peaks can 
be tuned with the rotational angle. These findings makes twisted bilayer blue phosphorene a strong candidate as a 
solar cell or photodetection device. Our results are based on ab initio calculations of several rotated blue 
phosphorene layers.
\end{abstract}

\begin{keyword}
blue phosphorene, bilayer, rotated, ab initio 



\end{keyword}

\end{frontmatter}


\section{Introduction}
Starting with the synthesis of graphene\cite{Novoselov666} a new era of two dimensional (2D) materials have emerged. 
Among them we find hexagonal boron nitride, molybdenum disulphide, dichalcogenides, layered 
oxides\cite{Novoselov26072005}, silicene and more recently one layer of black phosphorous 
(phosphorene)\cite{Phosphorene_2014}. 

These materials, in general, have diverse properties, but some of them are semiconductors, a dream of stacking them 
in a certain order has become a reality resulting in structures with desired properties\cite{Geim_2013}. It has been 
possible not only to stack them in certain order but also rotate layers as well to exploit certain properties that 
twisted structures posses\cite{Rotation_2016}. For instance in twisted bilayer graphene a plethora of interesting 
angle dependent properties have been found
\cite{Lopes_2007,Li_2010,Morell_2010,Moon_2013,Trambly_2012b,HeLin_2012,Morell_2014a,Morell_2015}.

Single layer phosphorene  is a direct band gap semiconductor (2 $eV$). The addition of more layers reduce the value 
of the gap, allowing in principle to control its size which goes inverse with the layer number down to the value of 
the bulk (0.3 $eV$)\cite{Phosphorene_2014,GapP_2014}. Nonetheless the main optical absorption peak of monolayer 
phosphorene is in 1.2 $eV$ due to the presence of excitons\cite{Tran2014}.

On the other hand it has been shown in Ref.[\cite{Tomanek_2014}] that a displacement of some atoms in crystal 
phosphorene results in a honeycomb structure like graphene. The authors studied its stability by first principle 
methods and found that it can be as stable as phosphorene; this new 2D allotrope, named blue phosphorene, has  an 
unit cell composed of two sublattices, like in honeycomb lattice but with the difference that the two atoms are not 
in the same horizontal plane, forming a periodically buckled topology. The  blue   phosphorene is   a   
semiconductor   with an indirect band gap of $\sim$ 2 $eV$.

Along  this  line a theoretical  work based on first 
principles calculation proposes the use of  blue phosphorene nanoribbons\cite{Xie2014} to modulate  the  band  gap, 
in this case the gap decreases for narrower nanoribbons. Nonetheless the system remains an indirect band gap 
semiconductor for either armchair or zigzag edges.

Another way to modulate  the physical  properties  of blue phosphorene is changing the  stacking. Two blue 
phosphorene layers stacked in AA or AB fashion have also  shown  a decrease  in  the  band  gap  compared   with  
the  blue   phosphorene  monolayer. Unfortunately the band gap  is also indirect\cite{BluePwithField_2015} which is 
a real limitation to use this 2D allotrope in the solar cell industry due to the poor efficiency that characterizes 
indirect band gap semiconductors\cite{Si_Directgap}. 

In this article we show that twisted bilayer blue phosphorene(TBBP) is a direct band gap semiconductor at least in 
three of the four structures studied. The relative rotation angle (RRA) between layers modulates the value of the 
gap. The band gap goes from 0.95 $eV$ for a RRA of 7.3$^\circ$ to a value of 1.6 $eV$ for an angle of 21.8$^\circ$. 
Unlike stacked AA and AB bilayer blue phosphorene, TBBP has absorption peaks in the visible range of the absorption 
spectra, and additionally the presence of flat valence bands in TBBP structures could improve the optical absorption 
probability. These findings makes TBBP a good candidate for a solar cell device, since an ideal solar cell device 
would be composed of a battery of stacked direct band gap semiconductors, each of them with a different value of the 
band gap and hopefully around 1.5 $eV$ to take advantage of the solar spectrum\cite{Zdanowicz2005757,Dai_2014}. 

\begin{figure}[t]
   \centering
\includegraphics[width=1.0\linewidth]{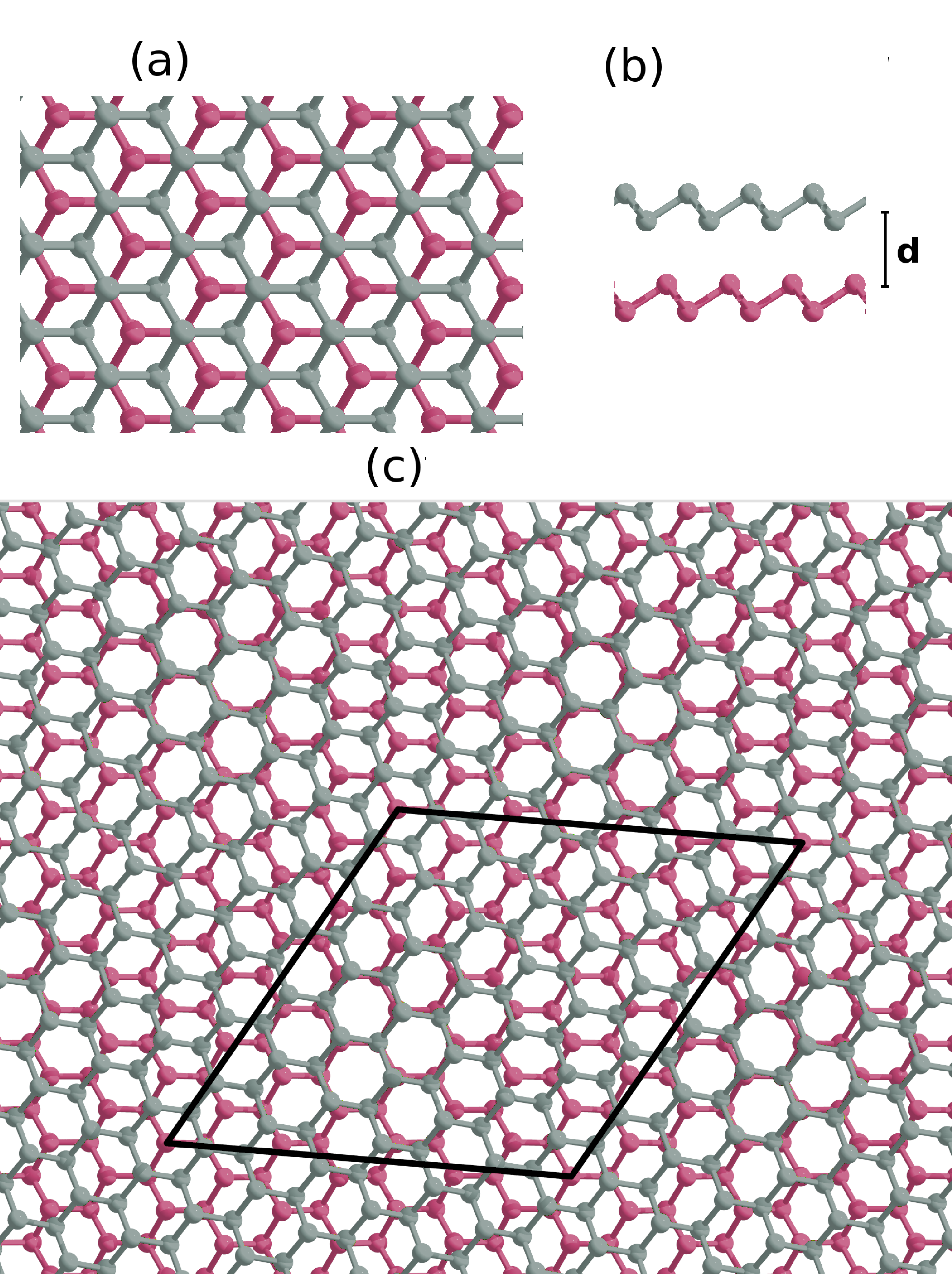}
\caption{\label{xyz}  (a) Two stacked AB blue phosphorene layers.
(b) Lateral view of the two layers in AB stacking. (c) Twisted structure for a relative rotating angle of  $9.34^{\rm o}$.}

\end{figure}

\section{Method}

\subsection{Geometry of the Unit cell}

Crystalline blue phosphorene is composed of an hexagonal lattice with a basis of two P atoms like the honeycomb 
structure of graphene. The results of relaxation processes determined by \textit{ab initio} methods shows that the 
two atoms are not in the same plane. In fact, there is a vertical shift between them of 1.24 \,\AA 
\cite{BluePwithField_2015}. On the other hand, one of the ways of stacking two layers is in the Bernal or AB 
stacking. In bilayer graphene it is the most stable configuration. The unit cell of stacked AB blue phosphorene will 
have four atoms, two in each layer. One of the atoms at the top layer will have directly below it an atom   of the 
bottom layer whereas the second atom of the upper layer will be located exactly above the center of a hexagon of the 
layer beneath, as shown  in Fig \ref{xyz} (a).

We consider a system composed of two coupled blue phosphorene layers when there is a RRA between them. The unit cell 
of the rotated structure was constructed following a procedure exposed in several previous
works\cite{Morell_2010,Morell_2011b}. Starting from a stacked AB bilayer one rotates one layer from a point in the 
lattice to an equivalent location. This can be done by taking a vector $\vec{r}=m \vec{a}_{1} + n \vec{a}_{2} $ and 
rotating it around a line perpendicular to the layers passing through the origin to an equivalent position at  
$\vec{t}=n \vec{a}_{1} + m \vec{a}_{2}$, where $\vec{a}_{1}$ and
$\vec{a}_{2}$ are the monolayer lattice vectors; \textit{n},\textit{m} are integers. This way the rotational angle 
and all the relevant geometric information of the bilayer twisted cell can be expressed in terms of $n$ and $m$.

In Figure \ref{xyz}(a) we show the stacked AB blue phosphorene, in panel (b) a lateral view of the two layers where 
the buckled structure can be observed, and in panel (c) the twisted unit cell obtained doing a commensurate 
rotation. In this case we selected a hollow site of the stacked AB bilayer as a center of rotation, which is also 
the origin of the twisted unit cell. The hollow site is the point where an atom of the bottom layer have directly 
above a center of an hexagon of the top layer.    

\begin{figure}[h]
     \centering
     \includegraphics[width=1.0\linewidth]{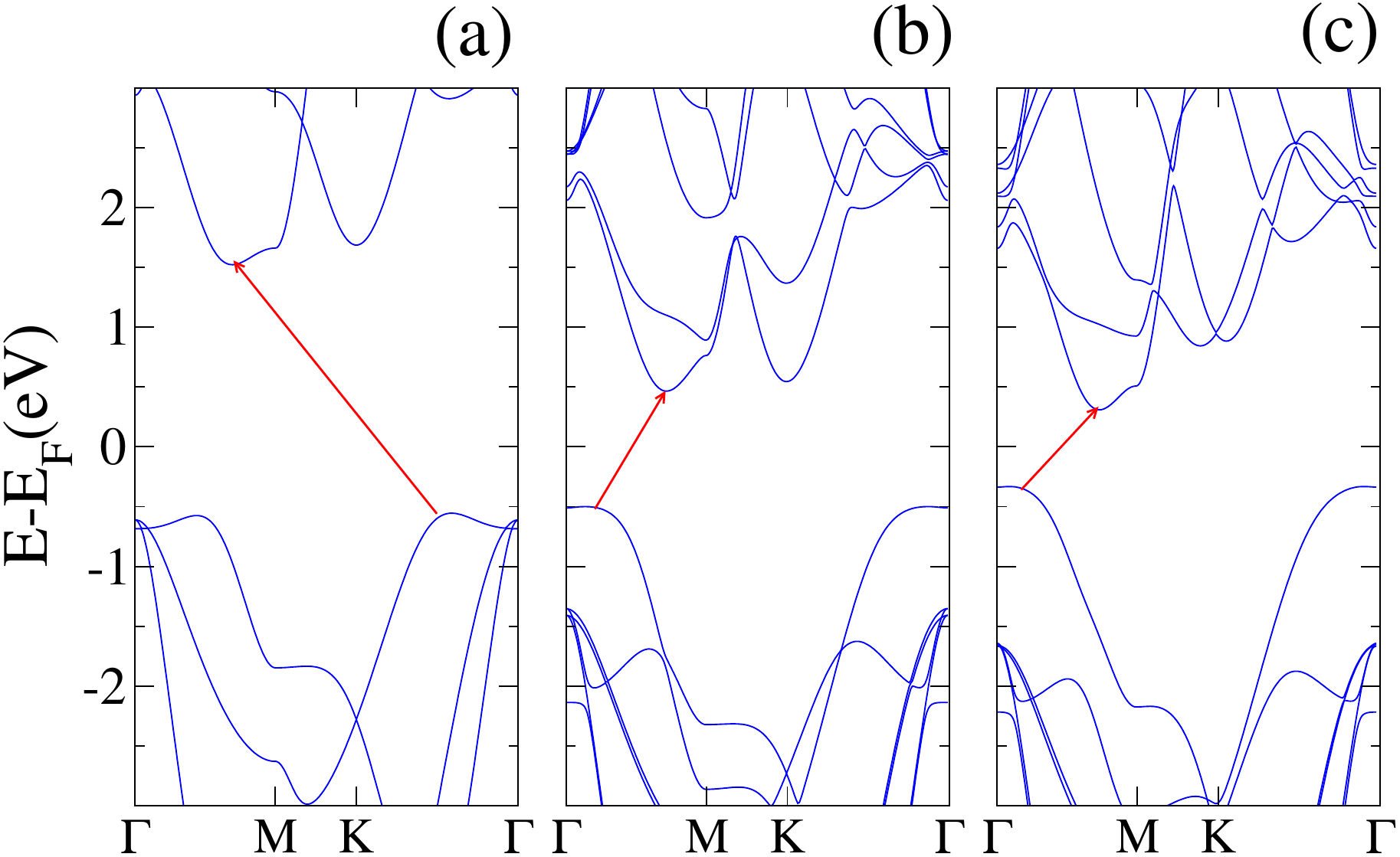}
        \caption{Bands  structure for three different  blue  phosphorene  structures (a)  monolayer, (b) stacking AA       
        and (c) stacking AB. The red arrow shows the indirect band gap.}\label{f1}
\end{figure}

\subsection{DFT calculations}

The calculations were performed within the density functional theory (DFT) approach, as implemented in the SIESTA 
code\cite{siesta}.  The Kohn-Sham orbitals were expanded in a localized orbitals basis set (double-$\xi$, single 
polarized in the present work). The twisted blue phosphorene bilayers were simulated considering cells with $28$ 
(TBBP-28), $76$ (TBBP-76),  $148$ (TBBP-148)  and $244$ (TBBP-244) atoms,  corresponding to RRA ($\theta$)  of 
$21.8^\circ$, $13.2^\circ$, $9.4^\circ$, and  $7.3^\circ$ respectively, and with a vacuum region of $20$~\AA~. 
For the  exchange-correlation potential we   employed vdw-KBM approach\cite{Klimes2010}, that includes the van der 
Waals (vdW) interactions within the self-consistent process. All structures  were fully  relaxed  considering a 
force convergence tolerance of $0.01$~eV/\AA~ per atom. The Brillouin zone (BZ)  was sampled by a $3\times 3\times 
1$ Monkhorst-Pack mesh that ensure the total energy convergence. The optical absorption spectrum is obtained through 
the imaginary part of the dielectric function, implemented  in the SIESTA code using the  dipolar  approximation. 
This approximation does not  take into account the many  body effects as GW  corrections  and the  electron-hole  
interaction. These calculations are  prohibitive  for  such a big  systems   as  TBBPs. So  that, the  dipolar  
approach is certainly a  good first  order approximation to  understand the  physical properties   of  TBBPs,  in  
particular  the  optical  absorption spectrum. All  optical  spectra  were calculated with  a Gaussian 
smearing of  0.06~$eV$. To  ensure   the   convergence  of  the  optical  spectrum  and  the  total  density  
of  states   we  employed a k-point grids  of $51\times 51\times 1$ for AA and AB stacking, $20\times 20 \times 1$ 
for RRAs of $21.8^\circ$ and $13.2^\circ$,  and  $11\times 11 \times 1$ for RRAs  of $9.4^\circ$ and $7.3^\circ$ 
respectively.

\section{Results}

We have calculated the geometrical structure, the band structure,  the total density  of   states,  and  the  
optical  absorption   of monolayer BP, bilayers stacked AA and AB  of BP and TBBP  with  different RRA. The  main  
geometric  parameters  and the  band  gap  for   each system are  shown  in the  table~\ref{Tabla}. For  the  BP  
monolayer  the  results   obtained   by  us  for  the lattice   constant,  and   the band  gap  are   consistent   
with  a   previous  SIESTA  calculations\cite{Tomanek_2014}. In  the  case of BP stacked AA and AB  the results are 
in  agreement with  a  previous calculation\cite{BluePwithField_2015}, where PBE and DFT-D2  corrections were 
employed to  incorporate  a vdW  interaction. There, the authors  obtained  a  distance of $3.23$ and $3.20$~\AA~ 
for  AA and AB stacking, respectively. Our  results   for  the  binding  energy  of  all structures   show  that  
the  AA  staking is   more  favourable than  the  AB  and TBBP  with  a   binding   energy  per  atom  of  
$-0.027$~eV. In the case of TBBP, the RRA does not make any perceptible  variation in the binding energy 
per  atom, the value obtained for all  structures  were  $-0.022$~eV/atom. This  results   suggest   that   the 
TBBPs  are  more stable  than  the  AB  staking regardless of the RRA.   

\begin{table}[t]
 \begin{center}
\begin{tabular}{cccccc}
\hline
System     &              $ a$(\AA)     &   Gap (eV)     & d(\AA)       & $E_b$(eV/atom) \\ 
\hline
Monolayer  &              $3.37$        &   $2.098$      &              &          \\ 
Bilayer AA &              $3.38$        &   $0.976$      & $3.185$      & $-0.027$ \\ 
Bilayer AB &              $3.41$        &   $0.645$      & $2.993$      & $-0.017$  \\ 
Twisted $21.8^\circ$ &    $8.91$        &   $1.330$      & $3.188$      & $-0.021$  \\
Twisted $13.2^\circ$ &    $14.671$      &   $1.183$      & $2.996$      & $-0.022$  \\
Twisted $9.4^\circ$  &    $20.489$      &   $1.068$      & $2.917$      & $-0.022$   \\
Twisted $7.3^\circ$  &    $26.320$      &   $0.949$      & $2.933$      & $-0.022$  \\
\hline
\end{tabular}
\end{center}
\caption{Geometric  parameters, band gap, and binding  energy  of several blue phosphorene  configurations. $a$ - lattice  constant, $d$ - maximal  distance  between layers and $E_{b}$ - binding  energy  by  atom.}
\label{Tabla}
\end{table}

In Figure \ref{f1} we show the results of the DFT calculations of the band structure for  monolayer and  bilayer 
staked AA and AB of  blue   phosphorene, the red arrow indicates the indirect band gap. The three structures are 
indirect band gap semiconductors, the size of the gap varies from $\sim$ 0.6 to 2 $eV$.

\begin{figure}[t!]
     \centering
     \includegraphics[width=1.0\linewidth]{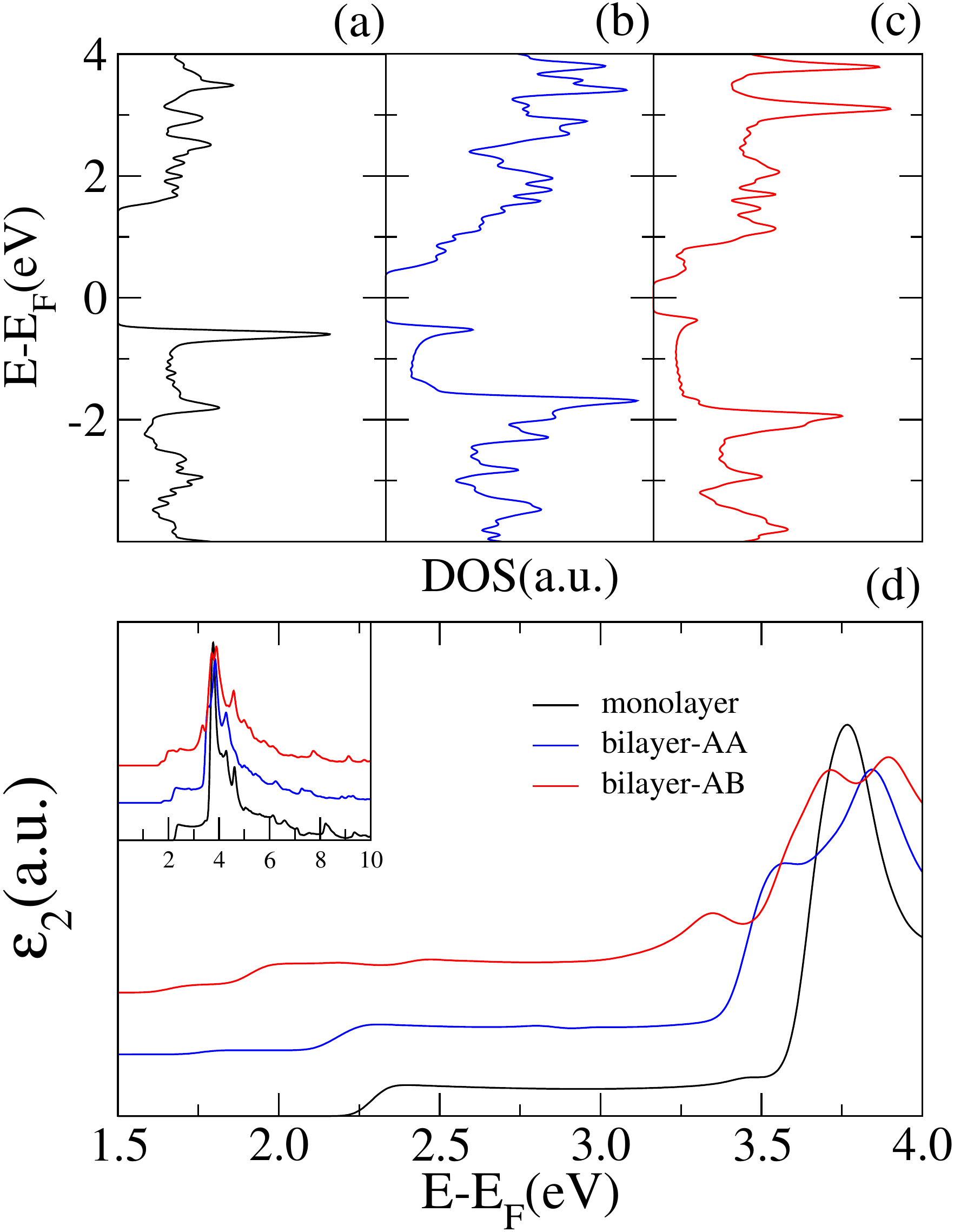}
        \caption{Total  density  of  states   for three different  blue  phosphorene  structures (a)   monolayer, 
        (b) stacking AA  and (c) stacking AB. Panel  (d) is for the imaginary  part  of the dielectric 
        function for monolayer,  staking AA,  and staking AB  of blue  phosphorene, the insert in the  panel (d) 
        shows the  imaginary part   of the  dielectric  function  in  different  incident   photon  energy  
        range.}\label{f2}
\end{figure}

The density of states (DOS) of the same three structures (Figure \ref{f2}(a)) shows clearly the evolution of the 
size of the band gap. The top valence band in the two bilayer systems is almost flat around the $\Gamma$ point of 
the BZ, and the vHs below the Fermi level can be associated with this valence band. A strong absorption of a 
material at a given energy is usually associated with the presence of vHs in the DOS. The absorption spectra in 
Figure \ref{f2}(b)) show strong absorptions peaks for energies around $\sim$ 4 $eV$, but below this energy and above 
the absorption edges there is not any strong absorption peak whatsoever. This fact is certainly similar to what is 
observed   in other hexagonal  flat   structures   as   graphene   which  is transparent   in the  visible   
spectrum  range.

\begin{figure}[t!]
     \centering
     \includegraphics[width=1.0\linewidth]{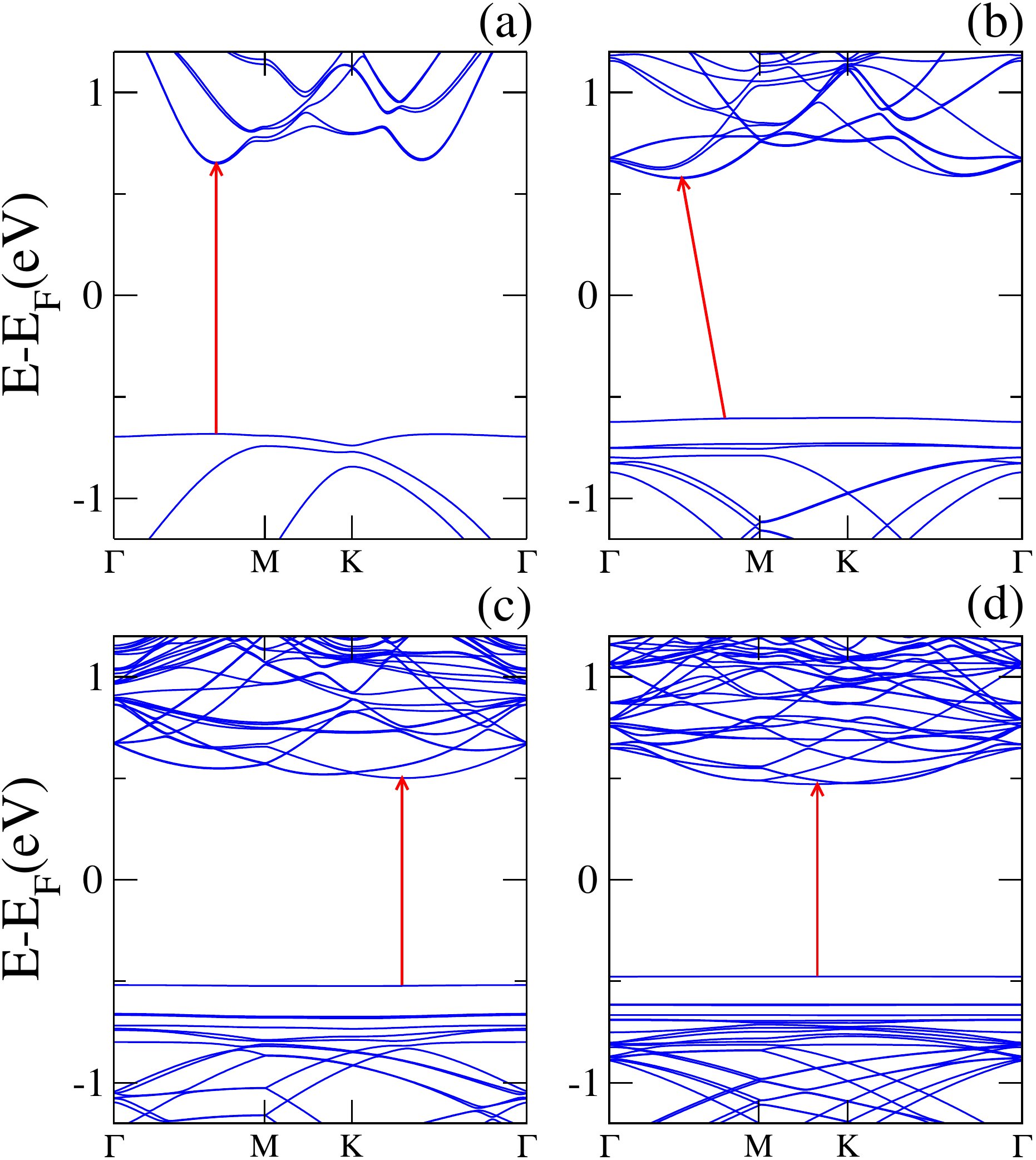}
        \caption{Bands  structure of four twisted bilayer blue phosphorene structures, in each  panel a different rotating angle (a)  
        $\theta$=21.8$^\circ$,  (b) $\theta$=13.2$^\circ$, (c) $\theta$=9.4$^\circ$ and (d) $\theta$=7.3$^\circ$. The (red vertical) 
        arrows show the band gap.}\label{f3}
\end{figure}

We were interested to see what happens when we rotate the layers. According to our previous experience, many 
fascinating angle dependent properties may appear. We have calculated the band structure of four twisted structure 
by DFT, and the  results  are  shown  in the  Figure \ref{f3}. It is possible to observe that the rotation angle 
changes the electronic properties of the system. in fact, for certain values of the relative rotation angle, the 
bilayer blue phosphorene becomes a direct band gap semiconductor. At least in three of the four structures the 
minimum distance between top valence band an lower conduction band lies at the same k-point. However the top valence 
band is almost flat throughout the whole BZ and, as long as the angle is smaller this effect is reinforced. One 
might explain such a behaviour as a result of the shrinking of the BZ. The flat region around the $\Gamma$ point 
observed in stacked AB bilayer have become relatively larger in the reduced zone of the twisted structures. As a 
result of this band flatness, optical absorption peaks will appear near the band gap value with amplitudes larger 
than those appearing in the stacked AB and AA structures.

The value of the band gap changes with the angle. Likely, more results are needed in order to establish the full 
variation pattern but in accordance with the four cases investigated here it is possible to advance the decrease of 
the band gap as a function of the rotation angle. In table \ref{Tabla} the values of the band gap for the angles 
studied are listed. It has a value of 1.33 $eV$ for the larger angle, 21.8$^\circ$ and goes to 0.95 $eV$ for a RRA 
of 7.3$^\circ$. The distance between layers in the relaxed structures diminishes slightly as the angle gets lower. 
The latter is expected since our starting point was a stacked AB bilayer.

\begin{figure}[t]
     \centering
     \includegraphics[width=1.0\linewidth]{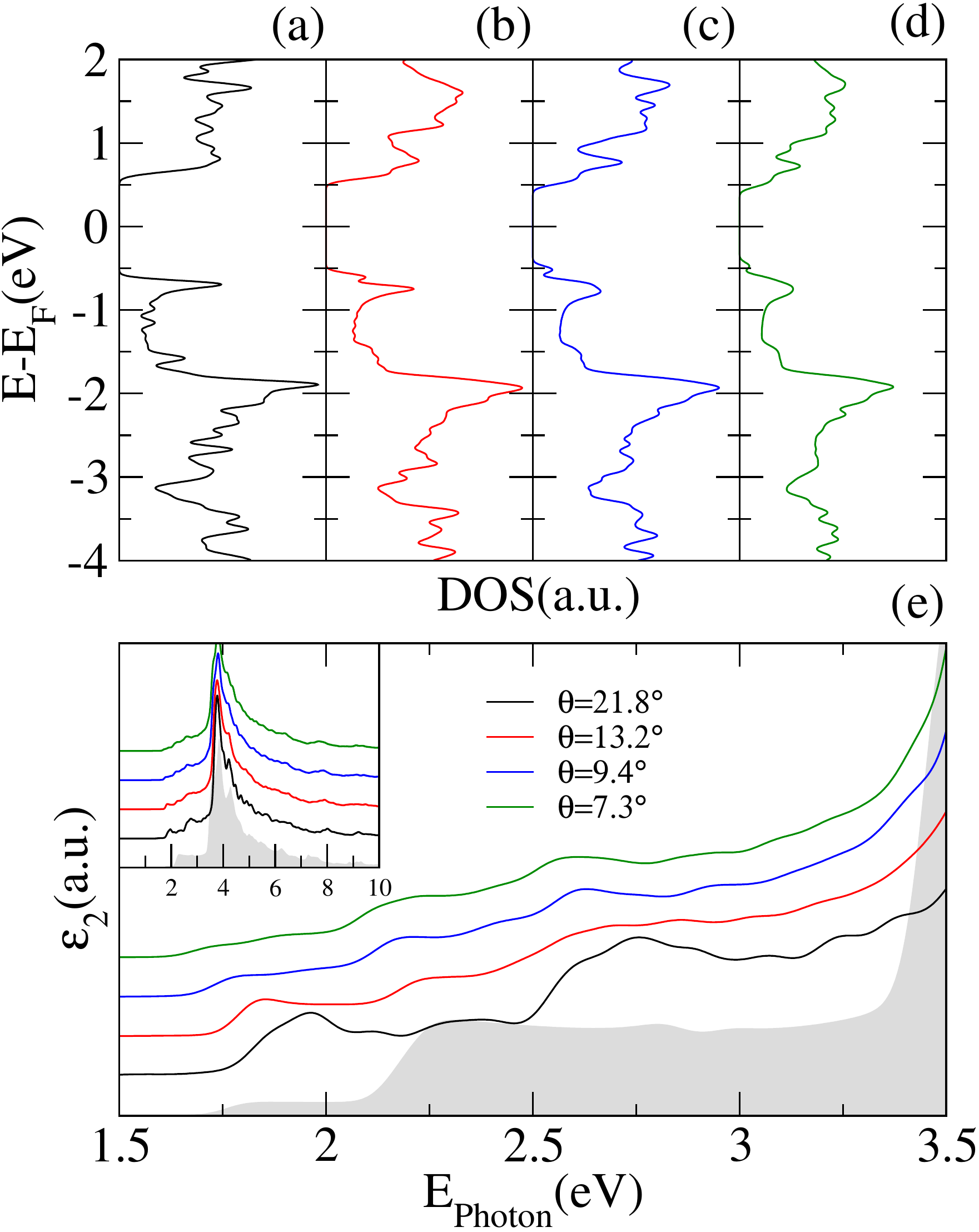}
        \caption{Total  density  of  states  of four twisted bilayer blue phosphorene structures, in each  panel a 
        different rotating angle (a) $\theta$=21.8$^\circ$,  (b) $\theta$=13.2$^\circ$, (c) $\theta$=9.4$^\circ$ and 
        (d) $\theta$=7.3$^\circ$. Panel  (e) is for the imaginary  part  of the dielectric of four twisted bilayer 
        blue phosphorene structures, the insert in the  panel (e) shows the  imaginary part   of the  dielectric  
        function  in  different  incident   photon  energy range.}\label{f4}
\end{figure}
  
The DOS of the twisted structures, Figure \ref{f4}(a), express in another format the facts we were mentioning; that 
is, the slight dependence of the band gap with the angle and the presence of vHs just in the band edge of the 
valence bands. The absorption spectra in Figure \ref{f4}(b) reveal the presence of absorption peaks below 2 $eV$, 
the position of the peaks changes with the angle and follows in principle the same behaviour of the band gap. The 
intensity of the peaks diminishes also for lower angles.

Its is important to notice that we have not included many-electron effects in our calculations. The presence of 
these flat bands which is in many cases related with the appearance of many body effects as well as the difference 
of the experimental optical band gap with the theoretical (single electron calculations) in its cousin bilayer black 
phosphorene lead us to suggest that further studies of these structure should be made considering many body effects. 
However the task is complicated due to the large size of the unit cell.

\section{Summary and conclusions}

We have studied by DFT calculations the electronic structure of two coupled layers of blue phosphorene and found 
that a relative rotation between layers changes the electronic properties of the system. Bilayer stacked AA or AB 
blue phosphorene are indirect band gap semiconductor while some of the twisted structures are direct band gap 
semiconductor and moreover the size of the band gap depends on the RRA. The absorption spectra show peaks in the 
visible range of the spectra which makes twisted bilayer blue phosphorene another candidate for solar cell 
applications. The dependence of the band gap with the angle suggests that stacking several of these structures 
together will increase the efficiency in solar cell devices as it will cover a wider range in energy of the solar 
spectrum as well as a larger interface.

\section*{Acknowledgement}

The authors acknowledge financial support from the Universidad de   Medell\'{i}n Research Office through Projects 
746 and 747 and  Chilean FONDECYT grant 11130129. DAO and CAD are grateful to the Colombian Agencies CODI-
Universidad de Antioquia (Estrategia de Sostenibilidad de la Universidad de Antioquia and project: "Manipulaci\'{o}n 
de propiedades optoelectr\'{o}nicas de nanoestructuras semiconductoras por ondas ac\'{u}sticas superficiales."), 
Facultad de Ciencias Exactas y Naturales-Universidad de Antioquia (CAD-exclusive dedication project 2015-2016), and 
El Patrimonio Aut\'{o}nomo Fondo Nacional de Financiamiento para la Ciencia, la Tecnolog\'{\i}a y la Innovaci\'{o}n, 
Francisco Jos\'{e} de Caldas.

\section*{References}

 \newpage



\end{document}